\documentclass[11pt]{article}
\usepackage{epstopdf}
\usepackage{amsmath,amssymb,amsthm}
\usepackage{algorithm}
\usepackage{algorithmic}
\newtheorem{Thm}{Theorem}
\usepackage{graphicx,subfigure}
\newcommand{\be}{\begin{eqnarray}}
\newcommand{\ee}{\end{eqnarray}}

\newcommand{\bk}{{\boldsymbol{k}}}

\newcommand{\br}{{\boldsymbol{r}}}

\newcommand{\bx}{{\boldsymbol{x}}}

\newcommand{\bL}{{\boldsymbol{L}}}

\newcommand{\bX}{{\boldsymbol{X}}}
\newcommand{\bY}{{\boldsymbol{Y}}}

\newcommand{\dff}{\stackrel{\triangle}{=}}
\usepackage{fullpage}

\title{Two Dimension Intensity Distribution of Ultraviolet Scattering Communication}
\author{Difan Zou~\thanks{Department of Computer Science,
  University of California,
  Los Angeles, CA, 90064, USA. knowzou@ucla.edu}~\footnote{This work was done when Difan Zou was a graduate student in USTC.}, Zhengyuan Xu~\thanks{Key Laboratory of Wireless-Optical Communications, Chinese Academy of Sciences, School of Information Science and Technology, University of Science and Technology of China, Hefei, Anhui 230027, China. xuzy@ustc.edu.cn}~ and Chen Gong~\thanks{Key Laboratory of Wireless-Optical Communications, Chinese Academy of Sciences, School of Information Science and Technology, University of Science and Technology of China, Hefei, Anhui 230027, China. cgong821@ustc.edu.cn}}

\date{}

\begin{document}
\maketitle
\begin{abstract}
Consider a ultraviolet (UV) scattering communication system where the position of the transmitter is fixed and the receiver can move around on the ground. To obtain the link gain effectively and economically, we propose an algorithm based on one-dimensional (1D) numerical integration and an off-line data library. Moreover, we analyze the 2D scattering intensity distributions for both LED and laser, and observe that the contours can be well fitted by elliptic models. The relationships between the characteristics of fitting ellipses and the source parameters are provided by numerical results.
\end{abstract}



\section{Introduction}
The non-line-of-sight (NLOS) ultraviolet (UV) communication serves as an alternative information transmission solution when the radio-frequency (RF) is prohibited, not only because the communication can still be maintained when the direct link cannot be guaranteed , but also the usage of UV waveband incurs extremely weak solar background radiation due to the existing UV solar blind waveband on the earth \cite{xu2008ultraviolet}. Recently, the issues of NLOS UV scattering communication are widely studied for the point to point (p2p) scenario, based on which the channel modeling \cite{Drost2013ultra}, signal characterization and performance analysis \cite{Drost2015deadtime,gong2015non,zou2018signal} are studied from both the transmitter and receiver sides. For UV communication networks, such as broadcasting and ad-hoc scenarios, the connectivity performance has been studied in \cite{wang2011connectivity,vavoulas2011connectivity}. However, the detail shape of coverage area and its relationships to the UV source parameters have not be characterized, which still need to be further investigated.  To achieve this goal, it is necessary to calculate the scattering intensity distribution of 2D receiver's positions.

Traditionally, the link gain of a scattering communication channel can be obtained based on the Monte Carlo method \cite{ding2009modeling,xu2015effects}, theoretical analysis \cite{xu2008analytical,wang2010non,zuo2013closed,sun2016closed} and experimental measurements \cite{liao2015uv, raptis2016power,wang2017demonstration}. However, large amount of link gains needs to be calculated to form a radiation distribution, and the Monte Carlo method may be intractable due to its huge time consumption. Similarly, the distribution is also impossible to be obtained by exhaustive testing on-site. Therefore, the analytical method may become the only approach that can be utilized to obtain the intensity distribution in an economical way. However, for a large field-of-view (FOV) of each receiver, the approximation on the scattering common volume in \cite{xu2008analytical,zuo2013closed} does not hold, thus a new theoretical approach needs to be provided.

In this paper, we allow the receiver to be able to capture photons as much as possible. In other words, the receiver FOV is assumed to be $\pi$ that each photon arriving at the receiving plane can be detected. Since the receiver has no information about the position of the transmitter, it may adjust the receiving plane such that the normal direction is perpendicular to the ground. Based on such assumptions, we first consider the laser source, and propose an analytical expression of link gain as a function of the elevation angle and the receiver's 2D position. To further reduce the computational complexity, an off-line link gain library is constructed and the link gain for given elevation angle and receiver position can be obtained via linear interpolation in the library. Moreover, we propose a detailed link gain library construction process and link gain calculation algorithm for an LED source typically with a large beam divergence angle. Based on such algorithm, we discuss the 2D scattering intensity distribution pattern, and observe that the contour could be well fitted by the elliptic model. Finally, the relationships between the characteristics of fitting ellipses and the UV source parameters (elevation angles and divergence angles) are investigated, where some numerical results are presented.

\section{Model and Algorithm}
In the following, we first present a novel algorithm to calculate the link gain in an economical way. Later on, the 2D scattering intensity distribution pattern discussions and studies on the relationship between contour shape of the distribution and LED source parameters are conducted.

In the atmosphere, a variety of phenomena may occur on the transmitted light, including the scattering and absorption. In general, the Rayleigh scattering coefficient $k_s^{ray}$, the Mie scattering coefficient $k_s^{mie}$, and the absorption coefficient $k_a$ are adopted to characterize the atmospheric scattering and absorption intensities. Moreover, let $k_s=k_s^{ray}+k_s^{mie}$ and $k_e=k_a+k_s$ denote the total scattering coefficient and total extinction coefficient, respectively.

In our proposed model, the transmitter's position is fixed, and the receiver's position can be anywhere on the ground. Hence, the two dimension (2D) intensity distribution of scattering radiation will be discussed in order to analyze the optimal position of the base station for the UV scattering broadcast network. To put it simply, it is necessary to analyze the contour of such 2D intensity distribution to determine the maximum communication area for different user's positions.

\begin{figure}[t!]
\centering
\includegraphics[width = 0.6\columnwidth]{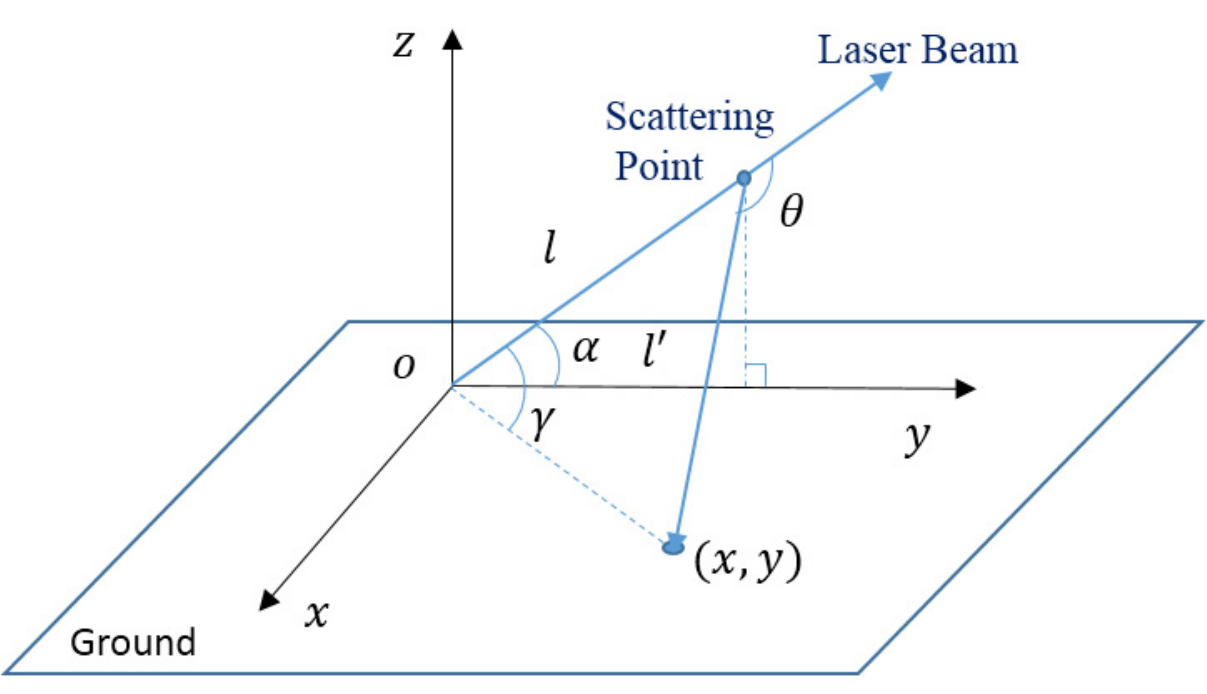}
\caption{The geometric description of the UV scattering radiation. }
\label{fig.geometry1}
\end{figure}

As presented in Figure 1, the transmitter's position is set to be $(0,0,0)$. Let $(x,y,0)$ denote the position of the receiver, $l$ and $l'$ denote the distance from scattering point to the transmitter and receiver, respectively. Let $\alpha$ be the elevation angle between the laser axis and the $Y$-axis, $\theta$ be the scattering angle between the light direction and the line from scattering point to the receiver, $\Omega(l)$ be the solid angle of the receiver at each scattering point $(0, l\cos\alpha,l\sin\alpha)$, and $A_r$ be the area of receiver aperture.

Let $E_t$ be the intensity of transmitted UV signal. The intensity of signal that scattered at $(0,l \cos \alpha, l \sin \alpha)$ and detected by the receiver is given by
\begin{equation}
\delta E_r\left(l\right) = E_t P\left[\mu(l)\right] \Omega(l) e^{-k_a\left(l+l'\right)}e^{-k_sl'} \delta l,
\end{equation}
where $\mu(l)$ is the cosine of scattering angle $\theta_s$, and a function with respect to $x$, $y$, $\alpha$, and $l$, given by $\mu(x,y,\alpha,l)=\frac{y\cos\alpha-l}{l'}$,
and $\Omega(l)$ denotes the solid angle from receiving area to the scattering point $(0,l \cos \alpha, l \sin \alpha)$, which is given by $
\Omega(x,y,\alpha,l)=\frac{A_r}{l'^2}\frac{l\sin\alpha}{l'}$.
$P\left(\mu\right)$ denotes the scattering phase function, which could be obtained from \cite{ding2009modeling}.
In the atmosphere, the free distance $l$ satisfies the exponential distribution $
f(l)=k_s e^{-k_s l}$ .
Then the total received power at $(x,y,0)$ is given by
\be\label{eq.linkgain_1d}
E_r&=&\int_{0}^{\infty} E_t P\left[\mu(l)\right] \Omega(l) e^{-k_a\left(l+l'\right)}e^{-k_sl'} k_s e^{-k_s l}dl \nonumber \\
&=&\int_{0}^{\infty} E_t P\left[\mu(l)\right] \Omega(l) k_s e^{-k_e\left(l+l'\right)}{\rm d}l,
\ee
where the parameters $x,y,\alpha$ are suppressed for shorter notations.

Hence, the link gain function $L_g(x,y,\alpha)$, as the ratio of $E_r$ over $E_t$, is given by
\be\label{eq.linkgain_1d}
L_g(x,y,\alpha)=\int_{0}^{\infty} P\left(\mu\right) \Omega(l) k_s e^{-k_e\left(l+l'\right)}{\rm d}l.
\ee

The 1D numerical integration can reduce the simulation time dramatically compared with the Monte Carlo method. This is mainly attributed to the very narrow beam of the source. However, it still may cost significant time when we need more than $10^5$ link gains to form a 2D intensity distribution. Our idea is to construct a link gain library that stores much data calculated according to (\ref{eq.linkgain_1d}), then the required link gain can be obtained via interpolation based on the data in such library. It can be seen that the library is three-dimensional, which contains the variables of $x$, $y$ and $\alpha$. However, in order to obtain a accurate and comprehensive library, there are still many link gains to be calculated, typically more than $10^7$, which would cost more than hundreds of hours to generate such library.

To address this issue, we aim to figure out if a library with a lower dimension could work. Motivated by this, we first provide the following result.

\begin{Thm}
If the link gain is calculated in accordance with (\ref{eq.linkgain_1d}), we have the following,
\be\label{eq.mapping2standard}
L_g(x,y,\alpha)=\frac{L_g(0,\sqrt{x^2+y^2},\beta)\sin \alpha}{\sin \beta}.
\ee
where $\beta = \arccos (\cos g \cos a)$ and $\gamma= \arctan x/y$.
\begin{proof}	
For the link gain $L_g(x,y,\alpha)$, we have
\be
l'^2&=&x^2+(y-l\cos\alpha)^2+l^2\sin^2\alpha \nonumber \\
&=&x^2+y^2+l^2-2yl\cos\alpha \nonumber \\
&=&y'^2+l^2-2y'l\cos\alpha',
\ee
where $y'=\sqrt{x^2+y^2}$ and $\cos\alpha'=\frac{y}{y'}\cos\alpha=\cos\alpha\cos\gamma$. This equation implies that $l'(x,y,\alpha,l)=l'(0,y',\alpha',l)$.
For the cosine of scattering angle, we have
\be
\mu(x,y,\alpha,l)=\cos\theta_s=\frac{y\cos\alpha-l}{l'}=\frac{y'\cos\alpha'-l}{l'},
\ee
which implies that $\mu(x,y,\alpha,l)=\mu(0,y',\alpha',l)$.
As for the solid angle $\Omega(x,y,\alpha,l)$, we have
\be
\Omega(x,y,\alpha,l)=\frac{A_rl\sin\beta}{l'^3}\frac{\sin\alpha}{\sin\beta}=\Omega(0,y',\alpha',l)\frac{\sin\alpha}{\sin\beta}.
\ee
Substituting the above results into equation (\ref{eq.linkgain_1d}), we have
\be
L_g(x,y,\alpha)&=&\int_{0}^{\infty} P\left(\mu\right) \Omega(l) k_s e^{-k_e\left(l+l'\right)}{\rm d}l \nonumber \\
&=&\int_{0}^{\infty} P\left(\mu'\right) \Omega'(l)\frac{\sin\alpha}{\sin\beta} k_s e^{-k_e\left(l+l'\right)}{\rm d}l \nonumber \\
&=&L_g(0,y',\alpha'),
\ee
where $\mu'\dff\mu(0,y',\alpha',l)$ and $\Omega'(l)\dff\Omega(0,y',\alpha',l)$.

\end{proof}
\end{Thm}

Thus, the 2D library $\bL=\left\{L(0,r_i,{\alpha_i}), r_i\in\br, \alpha_i\in\boldsymbol{\alpha}\right\}$ is constructed, where $\br$ denotes the set of different communication ranges when the receiver is located on the $Y$-axis, and $\boldsymbol{\alpha}$ denotes the set of different elevation angles (typically we set $\br=[0:1:1000]$ and $\boldsymbol{\alpha}=\pi\times[0.005:0.005:1]$). Then the link gain $L_g(x,y,\alpha)$ could be calculated by obtaining the link gain $L_g(0,(x^2+y^2)^{1/2},\beta)$ from the library $\bL$, and multiplying by a coefficient ${\sin\alpha}/{\sin\beta}$, where $\tan\beta=x/y$. For a special case that $\alpha=90^\circ$, it could be found that $\beta=90^\circ$, and  $L_g(x,y,90^\circ)=L_g(0,\sqrt{(x^2+y^2)},90^\circ)$, which implies the contour of 2D scattering intensity distribution has a circular shape under this case.

For the beam produced by LED, we assume the uniform pattern. In order to calculate the link gain for the LED source, we generate narrow beams with the uniformly distributed random direction, and average the corresponding link gains. Assuming the center direction of light is $(0,\cos\alpha,\sin\alpha)$, and the full divergence angle is $\phi_d$. Let $\boldsymbol{\zeta'}=(\zeta_x',\zeta_y',\zeta_z')$ denote the random emitting direction, which can be generated according to the coordinate transformation, specified by
\be\label{eq.tranformation1}
\zeta_x'&=&\sin \theta'\sin\phi', \nonumber \\
\zeta_y'&=&-\sin\theta'\cos\phi'\sin\alpha +\cos\theta'\cos\alpha,\nonumber \\
\zeta_z'&=&\sin\theta'\cos\phi'\cos\alpha+\cos\theta'\sin\alpha,
\ee
where $\theta'$ and $\phi'$ are random angles that are generated by $\theta'=\arccos\left[1-\xi(1-\cos\frac{\phi_d}{2})\right]$, and $\phi'=2\pi\xi$, where $\xi$ denotes a uniformly distributed random variable between zero and one.

Let $L_g'(x,y,\boldsymbol{\zeta'})$ denote the link gain for the narrow beam with normalized direction $\boldsymbol{\zeta'}$, which can be mapped into standard form $L_g(x',y',\alpha')$ via coordinate transformation, given by
\be\label{eq.tranformation2}
x'&=&\frac{\zeta_y'}{\sqrt{1-\zeta_z'^2}}x-\frac{\zeta_x'}{\sqrt{1-\zeta_z^2}}y, \nonumber \\
y'&=&\frac{\zeta_x'}{\sqrt{1-\zeta_z'^2}}x+\frac{\zeta_y'}{\sqrt{1-\zeta_z^2}}y, \nonumber \\
\alpha'&=&\arccos(\zeta_z').
\ee
Hence, for each narrow beam, we can obtain the corresponding link gain based on library $\L$. By averaging link gains of many random beams, we have the following results for the LED UV source,
\be
L_g^{LED}(x,y,\alpha,\phi_d)=\frac{1}{N}\sum_{k=1}^NL_g'(x,y,\boldsymbol{\zeta'_k}).
\ee
The whole calculation process is summarized in Algorithm~$1$.
\begin{algorithm}[t]
\caption{\small Calculating scattering radiation at $(x,y)$}
\begin{algorithmic}[1]
\STATE \textbf{Input:} Receiver position $(x,y)$, source beam elevation angle $\alpha$, full divergence angle $\phi_d$, link gain library $\bL$, number of random directions $N$.
\STATE\textbf{for} $count=1:N$
\STATE\quad\textbf{Initialize:} Generate random direction according to (\ref{eq.tranformation1});
\STATE \quad Update the receiver's coordinates from $(x,y)$ to $(x',y')$ according to (\ref{eq.tranformation2}) ;
\STATE \quad Update the elevation angle from $\alpha$ to $\alpha'$ according to (\ref{eq.tranformation2});
\STATE \quad Transform the link gain $L_g(x',y',\alpha')$ into the standard form according to (\ref{eq.mapping2standard});
\STATE \quad Calculate the link gain via linear interpolation based on the library $\bL$;
\STATE \quad Store the link gain in the vector $\bL_g^{x,y}$;
\STATE \textbf{end for}
\STATE \textbf{Output:} The average of variables in vector $\bL_g^{x,y}$.
\label{alg_scheme1}
\end{algorithmic}
\end{algorithm}

\section{Numerical experiment}
\begin{figure}[t!]
\centering
\includegraphics[width = 0.7\columnwidth]{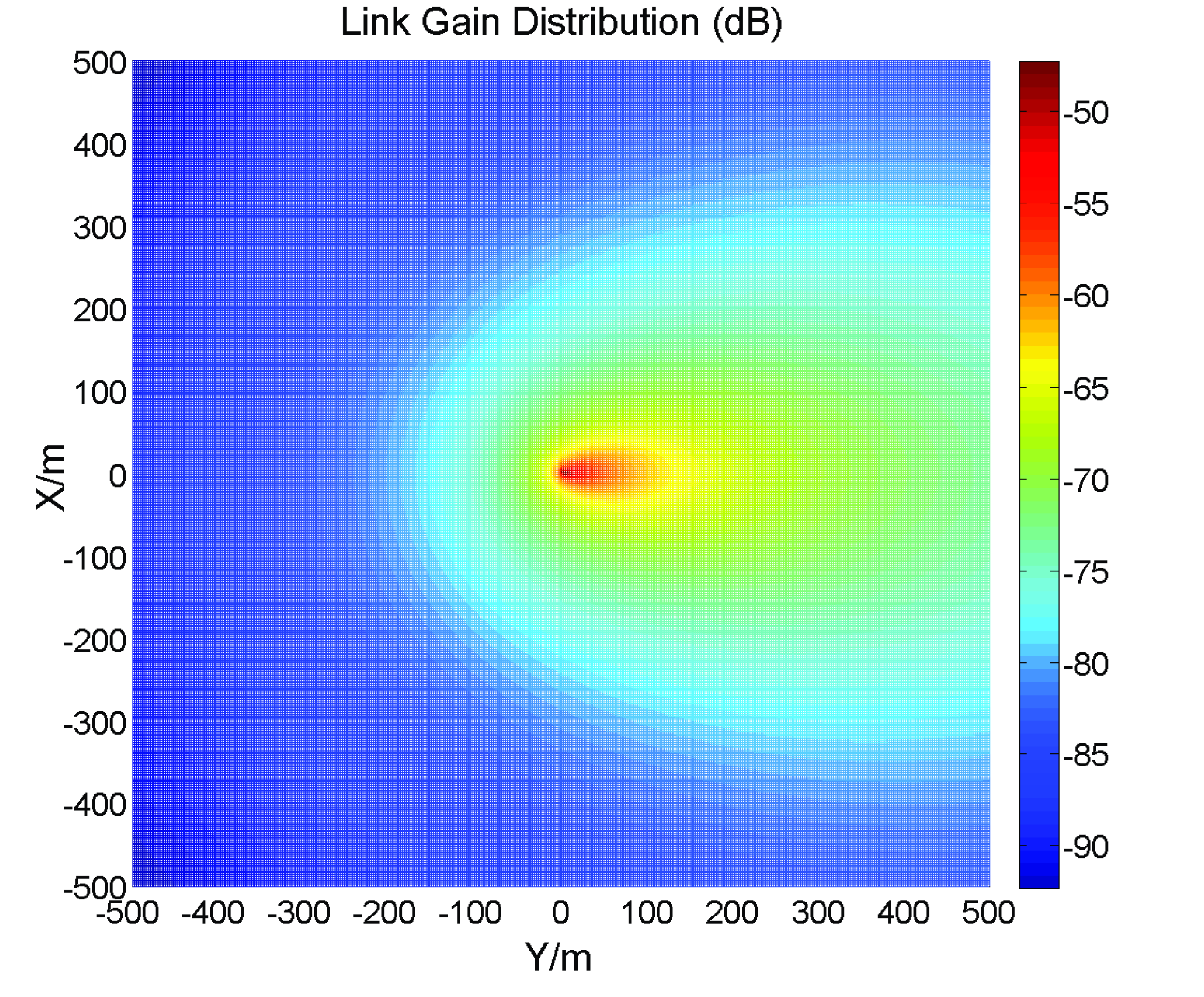}
\caption{The 2D radiation distribution for the laser source with elevation angle $\alpha=30^\circ$. }
\label{fig.dis1}
\end{figure}

Considering many possible receiver's positions, we obtain a 2D scattering intensity distribution in an economical way based on Algorithm 1. Specifically, assume the unit receiving area, we simulate the $500\times500$ link gains at the area $[-500,500]\times[-500,500]$$\rm m^2$ for each scattering intensity distribution. We first present the 2D radiation distribution for the laser source with elevation angle $\alpha=30^\circ$ in Figure~\ref{fig.dis1}.

\begin{figure}[htb]
	\centering
	\includegraphics[width = .8\columnwidth]{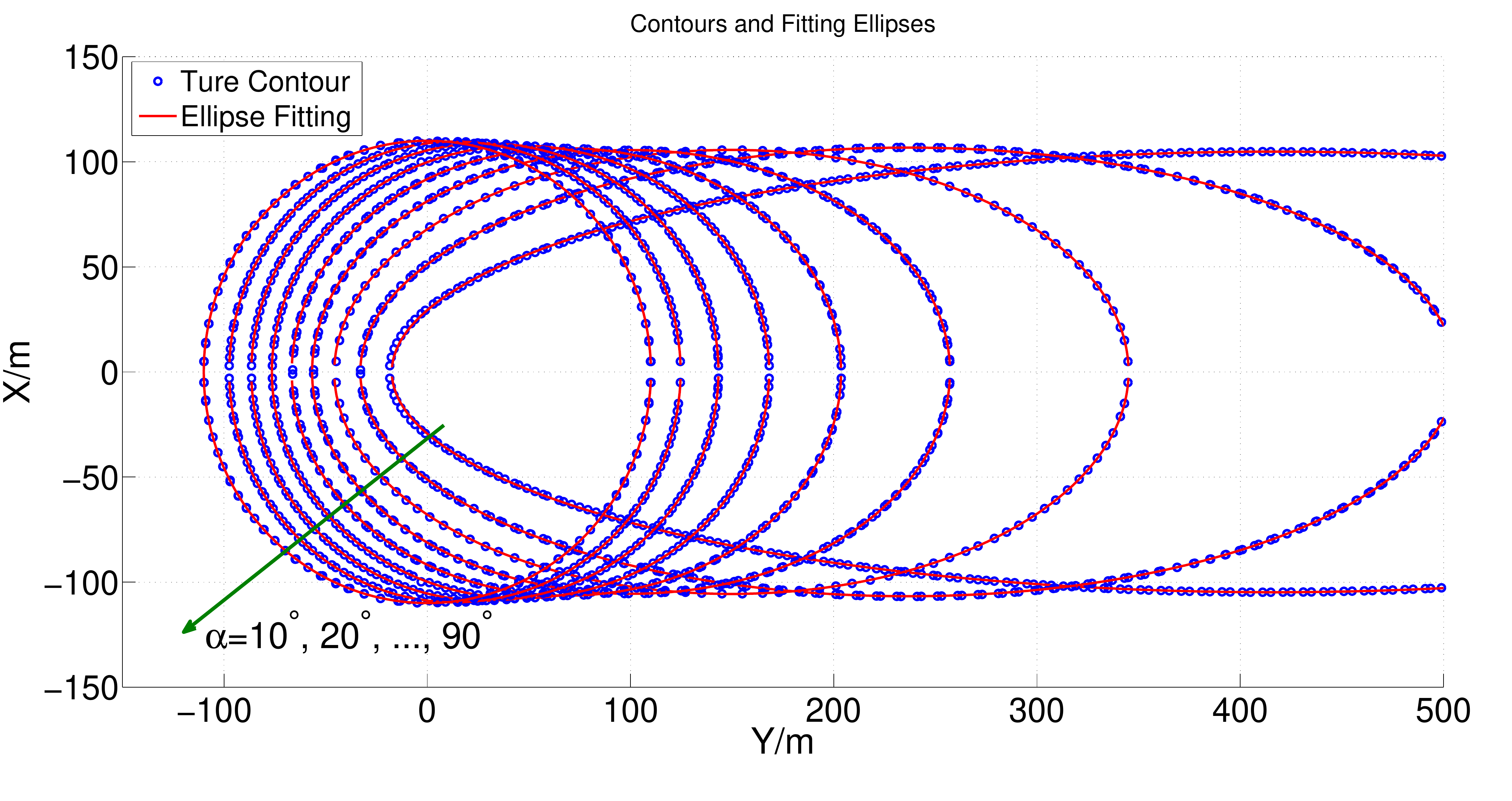}
	\caption{The contours and corresponding fitting ellipses of 2D radiation distribution for the laser source with elevation angle from $10^\circ$ to $90^\circ$ ($L_g=10^{-7}$). }
	\label{fig.contour1}
\end{figure}

It is clearly observed that the contour seems like a ellipse. Motivated by this, we extract the contour coordinates, denoted by $(\bX,\bY)=\{(x_i,y_i),i=1,2,\dots,M\}$. Then we perform an elliptic fitting on these data. Considering the elliptic function
\be
\frac{(x-x_0)^2}{a^2}+\frac{(y-y_0)^2}{b^2}=1,
\ee
where we have $x_0=0$ due to the fact $L_g(x,y,\alpha)=L_g(-x,y,\alpha)$.
In the fitting process, the ellipse parameters $y_0,a,b$ are derived based on the following least squares criterion
\be
\min_{y_0,a,b}\sum_{i=1}^M \left[a^2\left(1-\frac{y_0^2}{b^2}+\frac{2y_0}{b^2}y_i-\frac{y_i^2}{b^2}\right)-x_i^2\right]^2.
\ee
Let $\bk=\left[a(1-\frac{y_0^2}{b^2}),\frac{2a^2y_0}{b^2},-\frac{a^2}{b^2}\right]^T$ denote the parameter vector, whose optimal solution is given by
\be
k=(\bY_{M,3}^T\bY_{M,3})^{-1}\bY_{M,3}^T\bx_2,
\ee
where $\bY_{M,3}$ is a 3-order Vandermonde matrix of the parameters $\left\{y_1,y_2,\dots,y_M\right\}$
and $\bx_2\dff[x_1^2,x_2^2,\dots,x_M^2]^T$. Based on the estimated parameter vector $\bk$, the parameters $a$, $b$, and $x_0$ can be obtained according to the definition of vector $\bk$.

Figure~\ref{fig.contour1} shows the contours and corresponding fitting ellipses for the laser source with elevation angle from $10^\circ$ to $90^\circ$, where the link gain is set to be $L_g=10^{-7}$ . It can be seen that the ellipse curve can well model the contour of 2D radiation distribution, which verifies our hypothesis on the contour shape. When the elevation angle increases, the coverage area decreases and the contour becomes a circle when $\alpha=90^\circ$.

\begin{figure}[t!]
	\centering
	\subfigure[]{\includegraphics[width = 0.49\columnwidth]{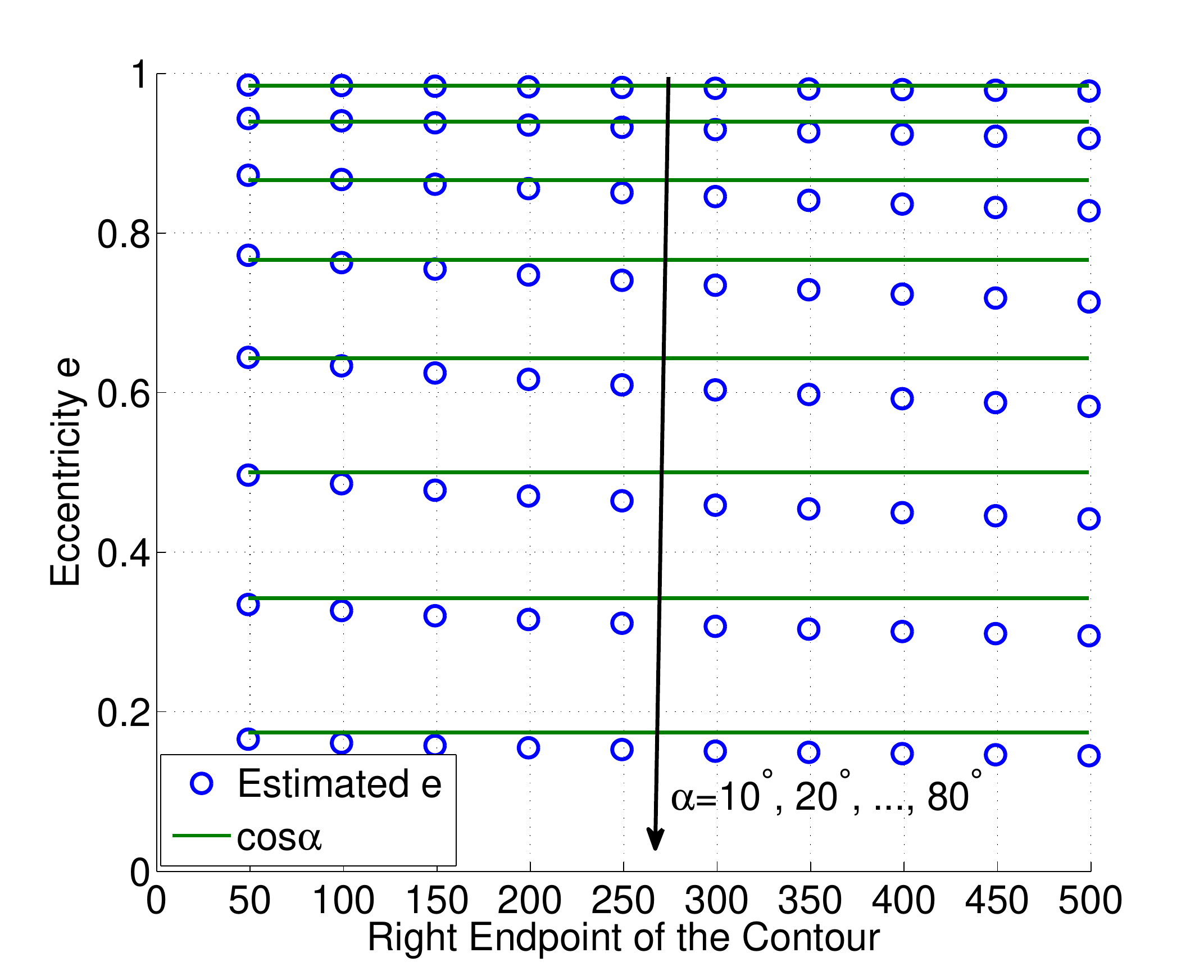}\label{fig.ecc1}}
	\subfigure[]{\includegraphics[width = 0.49\columnwidth]{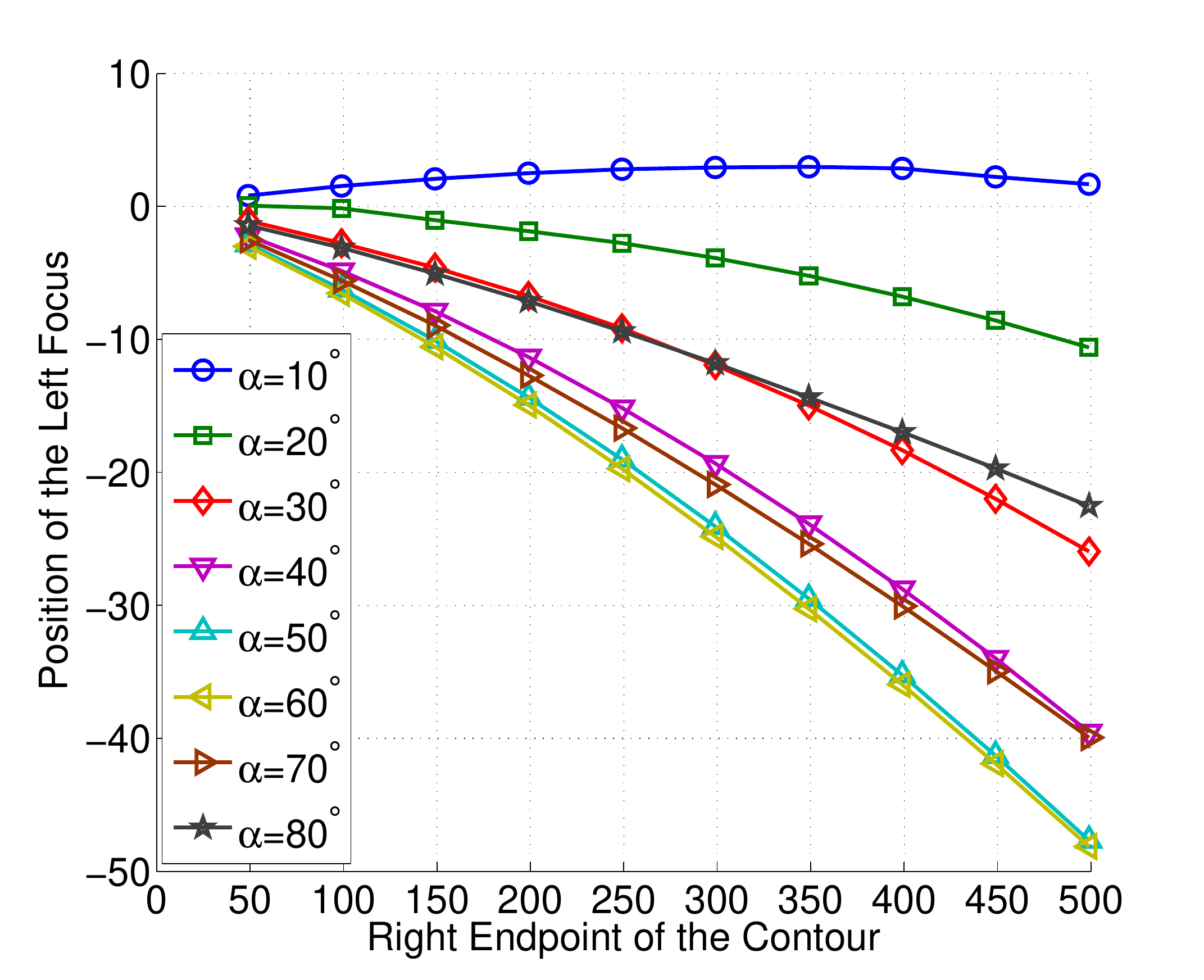}\label{fig.lfocus1}}
	\caption{(a) The eccentricity value with respect to the position of the right end point of the ellipse for different elevation angles. (b) The left focus position with respect to the position of the right end point of the ellipse for different elevation angles. }
\end{figure}

Considering the relationship between elliptic characteristics and the UV source parameters,  we first aim to figure out how the eccentricity values $e=\frac{\sqrt{b^2-a^2}}{b}$ of fitting ellipses vary with the elevation angle $\alpha$, where the contours at different positions are also considered. Figure~\ref{fig.ecc1} shows the relationship between the eccentricity and  contour position for different elevation angles. It can be seen that the eccentricity decreases with respect to the position of its right endpoint for all the elevation angles (except the case of $\alpha=90^\circ$). Meanwhile, we present the cosine of elevation $\alpha$ to fit the estimated eccentricity. It can be seen that such cosine fits well when the ellipse is small, and the gap between $\cos \alpha$ and the estimated result becomes large for a larger contour. Then it is natural to analyze the position of the fitting ellipses. Note that the 2D position of the transmitter, i.e. the coordinate $(0,0)$, is more likely a left focus. Thus we present the relation between the left focus position and the corresponding right endpoint in Figure~\ref{fig.lfocus1}. It is seen that the position of the transmitter is in fact not always a left focus, whose Y-axis coordinate decreases with respect to the size of contour when elevation angle $\alpha\ge20^\circ$. The case of $\alpha=10^\circ$ shows different characteristic that its Y-axis coordinate first increases and then decreases with respect to the ellipse size. We can also see that the gap between the transmitter position and left focus first increases and then decreases with respect to the elevation angle.

\begin{figure}[t!]
	\centering
	\includegraphics[width = 0.7\columnwidth]{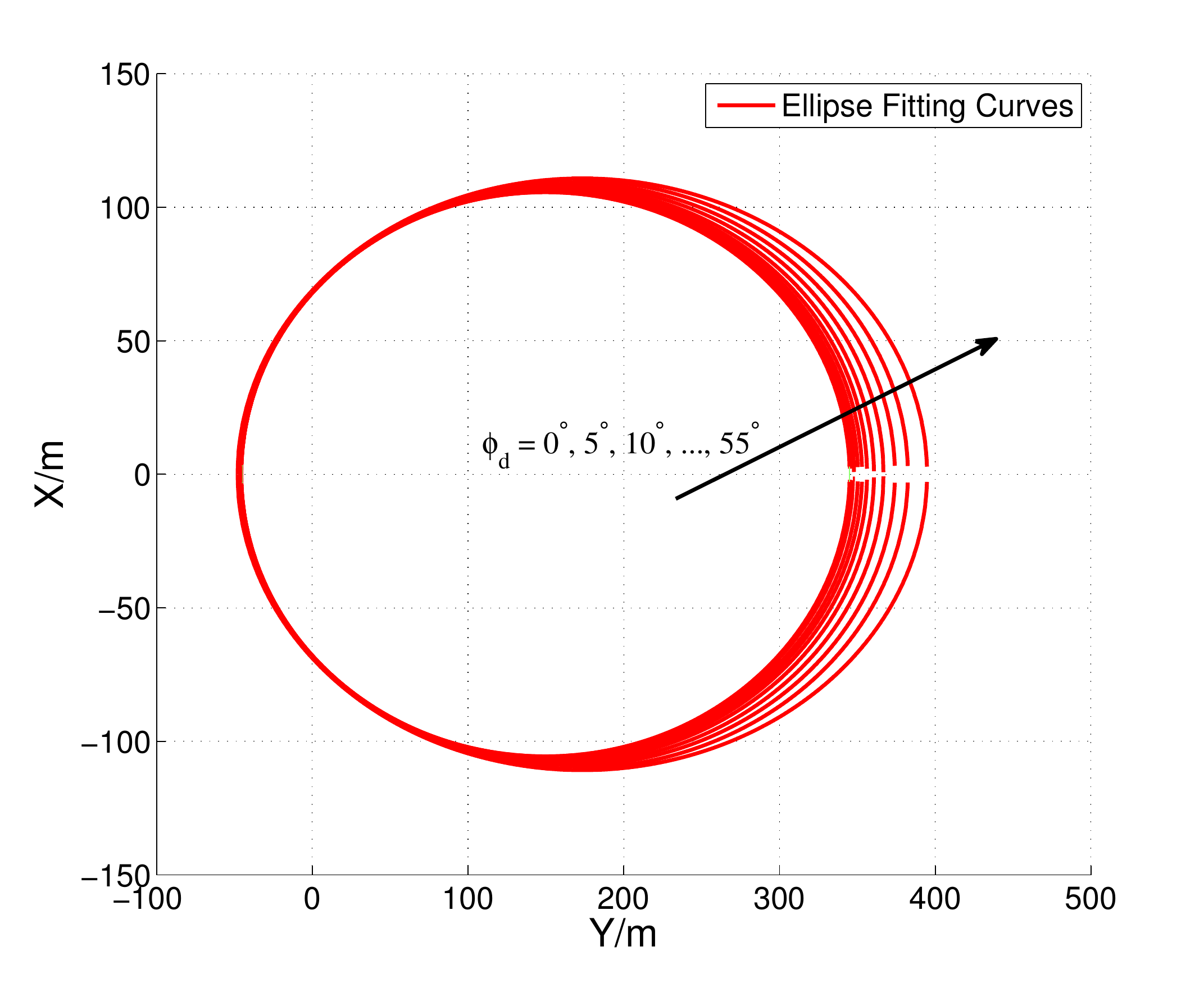}
	\caption{The ellipse fitting of the contour of 2D radiation distribution for the case $(\alpha,\phi_d)=(30^\circ,0^\circ-55^\circ)$ ($L_g=10^{-7}$). }
	\label{fig.contour2}
\end{figure}

Then we consider the relationship between the ellipse characteristics and the divergence angle $\phi_d$ when a large beam source such as LED is used. In Figure~\ref{fig.contour2}, it can be observed that the contour size increases with respect to the divergence angle when the corresponding link gain is fixed. However, the contour size is not very sensitive to the divergence angle. It can be seen that the left endpoints of the fitting ellipses basically remain the same and the right endpoints only vary from $(0,350)$m to $(0,400)$m.

\section{Conclusion}
In conclusion, we have proposed the basic concept of 2D scattering intensity distribution. In order to obtain large amount of link gains to form a 2D intensity distribution in an economical way, especially for the LED source, we have provided an novel algorithm based on a 2D link gain library.  Moreover, we have analyzed the geometry of 2D scattering intensity distribution and proposed an elliptic model to fit the contour. Then the relationship between the contour characteristics and the UV source parameters have been presented by numerical simulation. The results indicate that the contour size is not sensitive to the divergence angle.

\bibliographystyle{plain}
\bibliography{mybib}

\end{document}